\documentclass[11pt]{article}

\usepackage{amsmath}
\usepackage{amsthm}
\usepackage{amsfonts}
\usepackage{amssymb}
\usepackage{graphics}
\usepackage{graphicx}
\usepackage{amsbsy}

\newtheorem{theorem}{Theorem}[section]

\newtheorem{lemma}{Lemma}[section]
\numberwithin{equation}{section}
\DeclareMathAlphabet{\mathbi}{OML}{cmm}{b}{it} 
\newcommand{\bx}{\mathbi{x}}
\newcommand{\bk}{\mathbi{k}}
\newcommand{\bw}{\mbox{\boldmath$\omega$}}

\newcommand{\R}{\mathbb{R}}

\newcommand{\bdf}{\mathbi{f}}

\newcommand{\f}{\mathbi{f}}
\newcommand{\bu}{\mathbi{u}}
\newcommand{\bel}{\begin{equation}\label}
\newcommand{\ee}{\end{equation}}
\newcommand{\ben}{\begin{enumerate}}
\newcommand{\een}{\end{enumerate}}
\newcommand{\beq}{\begin{eqnarray}\label} 
\newcommand{\eeq}{\end{eqnarray}} 
\newcommand{\bc}{\begin{center}} 
\newcommand{\ec}{\end{center}} 
\newcommand\shalf{\ensuremath{{\scriptstyle\frac{1}{2}}}}
\newcommand\squart{\ensuremath{{\scriptstyle\frac{1}{4}}}}

\newcommand\Rey{\mbox{\textit{Re}}}  
\newcommand\bG{\mbox{\textit{Gr}}}  
\newcommand{\It}{\int_{t_{0}}^{t}}
\newcommand{\Idt}{\int_{\Delta t}}
\makeatletter
\@addtoreset{equation}{section}

\makeatother

\begin{document}
\bc
\textbf{\large Estimates for the two-dimensional Navier-Stokes equations in terms of the Reynolds number}
\ec
\bc
\textit{Invited paper for the JMP special issue on Mathematical Fluid Dynamics}
\ec
\bc
\textbf{\large J. D. Gibbon and G. A. Pavliotis}
\par\vspace{5mm}
Department of Mathematics,\\
Imperial College London, London SW7 2AZ, UK\,.
\ec
\begin{abstract}
The tradition in Navier-Stokes analysis of finding estimates in terms of the Grashof number 
$\bG$, whose character depends on the ratio of the forcing to the viscosity $\nu$, means that 
it is difficult to make comparisons with other results expressed in terms of Reynolds number 
$\Rey$, whose character depends on the fluid response to the forcing. The first task of this 
paper is to apply the approach of Doering and Foias \cite{DF} to the two-dimensional 
Navier-Stokes equations on a periodic domain $[0,L]^{2}$ by estimating quantities of physical 
relevance, particularly long-time averages $\left<\cdot\right>$, in terms of the Reynolds 
number $\Rey = U\ell/\nu$, where $U^{2}= L^{-2}\left<\|\bu\|_{2}^{2}\right>$ and $\ell$ is 
the forcing scale. In particular, the Constantin-Foias-Temam upper bound \cite{CFT} on the 
attractor dimension converts to $a_{\ell}^{2}\Rey\left(1 + \ln\Rey\right)^{1/3}$, while the 
estimate for the inverse Kraichnan length is $(a_{\ell}^{2}\Rey)^{1/2}$, where $a_{\ell}$ 
is the aspect ratio of the forcing. Other inverse length scales, based on time averages, and 
associated with higher derivatives, are estimated in a similar manner.  The second task is 
to address the issue of intermittency\,: it is shown how the time axis is broken up into 
very short intervals on which various quantities have lower bounds, larger than long 
time-averages, which are themselves interspersed by longer, more quiescent, intervals of time. 
\end{abstract}
%
%

\section{\large Introduction}

\subsection{General introduction}

In the last two decades the notion of global attractors in parabolic partial differential equations
has become a well-established concept \cite{CFT,CF,FMRT,IDDS}. The general nature of the dynamics 
on the attractor $\mathcal{A}$, in a time averaged sense, can roughly be captured by identifying 
sharp estimates of the Lyapunov (or fractal or Hausdorff) dimension of $\mathcal{A}$, or the 
number of determining modes \cite{JT}, with the number of degrees of freedom.  Introduced by 
Landau \cite{Landau}, this latter idea says that in a dynamical system of spatial dimension $d$ 
of scale $L$, the number of degrees of freedom  $\mathcal{N}$ is roughly defined to be that number 
of smallest eddies or features of scale $\lambda$ and volume $\lambda^{d}$ that fit into the 
system volume $L^d$
\bel{ndof1}
\mathcal{N} \sim \left(\frac{L}{\lambda}\right)^{d}\,.
\ee
This is the origin of the much-quoted $\mathcal{N}\sim \Rey^{9/4}$ result associated with the 
three-dimensional Navier-Stokes equations which rests on taking $\lambda \sim \lambda_{k}\sim
L\Rey^{-3/4}$, where $\lambda_{k}$ is the Kolmogorov length scale. In the absence of a proof of 
existence and uniqueness of solutions of the three-dimensional Navier-Stokes equations, at best 
this is no more than a rule of thumb result. It rests on a more solid and rigorous foundation, 
however, for the closely related three-dimensional LANS-$\alpha$ equations for which Foias, Holm 
and Titi \cite{FHT2} have proved existence and uniqueness of solutions. Following on from this, 
Gibbon and Holm \cite{GH06} have demonstrated that the dimension of the global attractor for this 
system has an upper bound proportional to $\Rey^{9/4}$. An important milestone has been passed 
recently in another closely related problem with the establishment by Cao and Titi \cite{CT} 
of an existence and uniqueness proof for Richardson's three-dimensional primitive equations 
for the atmosphere.

For the Navier-Stokes equations the idea sits more naturally in studies in the two-dimensional 
context. The existence and uniqueness of solutions has been a closed problem for many decades 
and the nature of the global attractor has been well-established [1-5,\,10-14].  While 
the two- and three-dimensional equations have the same velocity formulation, in reality, the 
former have a tenuous connection with the latter because of the absence of the drastic property 
of vortex stretching.  As a result, the presence of vortex stretching in three dimensions, and 
perhaps other more subtle properties, have set up seemingly unsurmountable hurdles even on periodic
boundary conditions. For problems on non-periodic boundaries, such as  lid-driven flow, solving 
the two-dimensional Navier-Stokes equations is a technically more demanding problem -- see some 
references in \cite{TemamSIAM,Glow2,Benartzi}.
\par\smallskip
The sharp estimate found by Constantin, Foias \& Temam \cite{CFT} for the Lyapunov dimension of 
the global attractor $\mathcal{A}$ expressed in terms of the Grashof number $\bG$
\bel{in1a}
d_{L}(\mathcal{A})\leq c_{1}\bG^{2/3}\left(1+ \ln\bG\right)^{1/3}\,,
\ee
has been one of the most significant results in two-dimensional Navier-Stokes analysis on a 
periodic domain $\Omega = [0,L]^{2}_{per}$.  The traditional length scale in the 
two-dimensional Navier-Stokes equations is the Kraichnan length, $\eta_{k}$, which plays an 
equivalent role in two dimensions to that of the Kolmogorov length, $\lambda_{k}$, which is 
more important in three dimensions. In two dimensions, $\eta_{k}$ and $\lambda_{k}$ are 
defined respectively in terms of the enstrophy and energy dissipation rates $\epsilon_{ens}$ 
and $\epsilon$
\bel{ensdef}
\epsilon_{ens}= \nu L^{-2}\left<\int_{\Omega}|\nabla\bw|^{2}\,dV\right>\,,
\hspace{2cm}
\epsilon = \nu L^{-2}\left<\int_{\Omega}|\bw|^{2}\,dV\right>\,,
\ee
where the pair of brackets $\left<\cdot\right>$ denote a long-time average defined as [2,3,10-13].
\bel{tadef}
\left<g(\cdot)\right> = \lim_{t\to\infty}\limsup_{g(0)}
\frac{1}{t}\int_{0}^{t}g(\tau)\,d\tau\,.
\ee
The inverse Kraichnan length $\eta_{k}^{-1}$ and the inverse Kolmogorov length 
$\lambda_{k}^{-1}$ are defined in terms of $\epsilon_{ens}$ and $\epsilon$ as
\bel{in1b}
\eta_{k}^{-1} = \left(\frac{\epsilon_{ens}}{\nu^3}\right)^{1/6}\,,
\hspace{2cm}
\lambda_{k}^{-1} = \left(\frac{\epsilon}{\nu^3}\right)^{1/4}\,.
\ee
It has been shown by Constantin, Foias and Temam \cite{CFT} that instead of using an estimate 
for $\epsilon_{ens}$ in terms of $\bG$, the upper bound for $d_{L}$ can be re-expressed in 
terms of $L\eta_{k}^{-1}$ (see other literature on this topic \cite{DG91,DGbook,G96})
\bel{in1c}
d_{L} \leq c_{2}\left(L\eta_{k}^{-1}\right)^{2}\left\{1+ \ln\left(L\eta_{k}^{-1}\right)\right\}^{1/3}\,.
\ee
If $d_{L}$ is identified with the number of degrees of freedom $\mathcal{N}$, this result is 
consistent with the idea expressed in (\ref{ndof1}) that in a two-dimensional domain, the 
average length scale of the smallest vortical feature $\lambda$ can be identified with the 
Kraichnan length $\eta_{k}$, to within log-corrections. The result in (\ref{in1a}) has also 
been improved by Foias, Jolly, Manley and Rosa \cite{FJMR2,FJMR1} to an estimate proportional 
to $\bG^{1/2}$ (to within logarithmic corrections) provided Kraichnan's theory of fully 
developed turbulence is implemented \cite{Kraichnan}. 
\par\smallskip
While these results display a pleasing convergence between rigorous estimates and scaling 
methods in the two-dimensional case, the tradition in Navier-Stokes analysis of finding estimates 
in terms of the Grashof number $\bG$, whose character depends on the ratio of the forcing to the 
viscosity $\nu$, means that it is difficult to compare with the results of scaling theories whose 
results are expressed in terms of Reynolds number. One of the tasks of this paper is to estimate 
quantities of physical relevance, particularly long-time averages, in terms of the Reynolds number,
whose character depends on the fluid response to the forcing, and which is intrinsically a property
of Navier-Stokes solutions. Doering and Foias \cite{DF} have addressed this problem and have shown 
that in the limit $\bG\to\infty$, solutions of the $d$-dimensional Navier-Stokes equations must 
satisfy\footnote{This result is not advertised in \cite{DF} but follows immediately from their 
equation (48).}
\bel{DF1}
\bG \leq c\,(\Rey^{2} + \Rey) \,,
\ee
while the energy dissipation rate $\epsilon$ has a lower bound proportional to $\bG$. The problem, 
however, is not as simple as replacing standard estimates in terms of $\bG$ by $\Rey^2$ from 
(\ref{DF1}). Estimates such as that for $d_{L}$ in (\ref{in1a}) and the inverse Kraichnan and 
Kolmogorov lengths defined in (\ref{in1b}), depend upon long 
time-averages of the enstrophy and energy dissipation rates defined in (\ref{ensdef}). Other 
estimates of inverse length scales (to be discussed in \S\ref{summary}) also depend upon long 
time-averages. When estimated in terms of $\Rey$ all these turn out to be better than straight 
substitution using (\ref{DF1}). These results are summarized in \S\ref{summary} and worked out 
in detail in \S\ref{2Dest}.
\par\smallskip
The second topic to be addressed in this paper is that of intermittency. Originally this 
important effect was considered to be a high Reynolds number phenomenon associated with 
three-dimensional Navier-Stokes flows. First discovered by Batchelor and Townsend \cite{BT49}, 
it manifests itself in violent fluctuations of very short duration in the energy dissipation 
rate $\epsilon$. These violent fluctuations away from the average are interspersed by quieter, 
longer periods in the dynamics. This is a well established, experimentally observable phenomenon 
\cite{Kuo71,MS91,Frischbk}; its appearance in systems other than the Navier-Stokes equations 
has been discussed in an early and easily accessible paper by Frisch \& Morf \cite{FM}. One 
symptom of its occurrence is the deviation of the `flatness' of a velocity signal (the ratio of 
the 4th order moment to the square of the 2nd order moment) from the value of 3 that holds 
for Gaussian statistics. 
\par\smallskip
Recent analysis discussing intermittency in three-dimensional Navier-Stokes flows shows that 
while it may be connected with loss of regularity, the two are subtly different issues \cite{GD05}.
This is reinforced by the fact that although solutions of the two-dimensional Navier-Stokes 
equations remain regular for arbitrarily long times, nevertheless many of its solutions at
high $\Rey$ are known to be intermittent \cite{inter1,inter2,inter3,inter4,inter5}. While 
three-dimensional analysis of the problem is based on the assumption that a solution exists 
\cite{GD05,DG02}, so that the higher norms can be differentiated, no such assumption is 
necessary in the two-dimensional case where existence and uniqueness are guaranteed. The result 
in both dimensions is such that the time-axis is broken up into good and bad intervals\,: on 
the latter there exist large lower bounds on certain quantities, necessarily resulting in their 
extreme narrowness and thus manifesting themselves as spikes in the data. This is summarized 
in \S\ref{summary} and worked out in detail in \S\ref{inter}.

\subsection{Summary and interpretation of results}\label{summary}

For simplicity the forcing $\bdf(\bx)$ in the two-dimensional Navier-Stokes equations 
($\hbox{div}\,\bu = 0$)
\bel{ns1}
\bu_{t}+\bu\cdot\nabla\bu = \nu\Delta\bu - \nabla p +\bdf(\bx)
\ee
is taken to be divergence-free and smooth of narrow-band type, with a characteristic single 
length-scale $\ell$ such that \cite{DF,GD05,DG02}
\bel{f1}
\|\nabla^{n}\bdf\|_{2} \approx \ell^{-n}\|\bdf\|_{2}\,. 
\ee
Moreover, the aspect ratio of the forcing length scale to the box scale is defined as 
\bel{aldef}
a_{\ell} = L/\ell\,.
\ee
With $f_{rms} = L^{-d/2}\|\bdf\|_{2}$, the usual definition of the Grashof number $\bG$ appearing 
in (\ref{DF1}) in $d$-dimensions is
\bel{Grdef}
\bG = \frac{\ell^{3}f_{rms}}{\nu^2}\,.
\ee
The Reynolds number $\Rey$ in (\ref{DF1}) is defined as
\bel{Redef}
\Rey = \frac{U\ell}{\nu}\,,
\hspace{3cm}
U^{2} = L^{-d}\left<\|\bu\|^{2}_{2}\right>\,,
\ee 
where $\left<\cdot\right>$ is the long-time average defined in (\ref{tadef}). One of the main 
results of this paper is the following theorem whose proof is given in \S\ref{theorem1proof}. 
All generic constants are designated as $c$.
\begin{theorem}\label{theorem1} 
Let $\bu(\bx,\,t)$ be a solution of the two-dimensional Navier-Stokes equations (\ref{ns1}) 
on a periodic domain $[0,\,L]^2$, and subject to smooth, divergence-free, narrow-band forcing 
$\bdf(\bx)$. Then estimates in terms of the Reynolds number $\Rey$ and the aspect ratio 
$a_{\ell}$ for the inverse Kraichnan length $\eta_{k}^{-1}$, the attractor dimension 
$d_{L}$, and the inverse Kolmogorov length $\lambda_{k}^{-1}$ are given by
\bel{en4b}
L\eta_{k}^{-1} \leq c\,(a_{\ell}^{2}\Rey)^{1/2}\,,
\ee
\bel{en4a}
d_{L} \leq c\,a_{\ell}^{2}\Rey\left[1 + \ln\Rey\right]^{1/3}\,,
\ee
\bel{en4c}
L\lambda_{k}^{-1} \leq c\,a_{\ell}\Rey^{5/8}\,.
\ee
\end{theorem}
\par\smallskip\noindent
In the short proof of this theorem in \S\ref{theorem1proof}, the estimate for $d_{L}$ in 
(\ref{en4a}) is not re-worked from first principles but is derived from a combination 
of (\ref{en4b}) and (\ref{en4a}). The result in (\ref{en4c}) comes from a $\Rey^{5/2}$ 
bound on $\left<H_{1}\right>$ and has also recently been found by Alexakis and Doering 
\cite{Alex}.  It implies that 
\bel{alex1}
\frac{L\epsilon}{U^3} \leq c\,a_{\ell}\Rey^{-1/2}\,,
\ee
whereas in three-dimensions the right hand side is $O(1)$. The estimate in (\ref{en4a}) is 
also consistent with the result of Foias, Jolly, Manley and Rosa \cite{FJMR2} when their 
$\bG^{1/2}$ estimate is converted to one proportioanl to $\Rey$. Their estimate, however, 
was based on the implementation of certain features of the Kraichnan model \cite{Kraichnan}, 
while (\ref{en4a}) is true for all solutions and requires no assumption of fully developed 
turbulence. 
\par\smallskip
The estimates for $\eta_{k}^{-1}$ and $d_{L}$ are consistent with the long-standing belief 
that $\Rey^{1/2}\times\Rey^{1/2}$ grid points are needed to numerically resolve a flow; 
indeed, when the aspect ratio is taken into account, Theorem \ref{theorem1} is consistent 
with $a_{\ell}\Rey^{1/2}\times a_{\ell}\Rey^{1/2}$. However, both these estimates are 
dependent upon only the time average of low moments of the velocity field.  For non-Gaussian 
flows, low-order moments are not sufficient to uniquely determine the statistics of a flow.  
Thus it is necessary to find ways of estimating small length scales associated with 
higher-order moments. In \S\ref{timeavs} we follow 
the way of defining inverse length scales associated with derivatives higher than two, 
introduced elsewhere \cite{DGbook}, by combining the forcing with higher derivatives of the 
velocity field such that
\bel{Fdef1a}
F_{n} = \int_{\Omega}\left( |\nabla^{n}\bu|^{2} + \tau^{2}|\nabla^{n}\bdf|^{2}\right)\,dV\,,
\ee
where $\tau = \ell^{2}\nu^{-1}[\bG(1+\ln\bG)]^{-\shalf}$ is a characteristic time\,: this 
choice of $\tau$ is discussed in Appendix \ref{App1.1}. The gradient symbol $\nabla^{n}$ 
within (\ref{Fdef1a}) refers to all derivatives of every component of $\bu$ of order $n$ 
in $L^{2}(\Omega)$. The $F_{n}$ are used to define a set of time-dependent inverse length 
scales
\bel{2nmom1}
\kappa_{n,r}(t) = \left(\frac{F_{n}}{F_{r}}\right)^{\frac{1}{2(n-r)}}\,.
\ee
Actually, $\kappa_{n,0}^{2n}$ behaves as the $2n$th-moment of the energy spectrum as shown by
\bel{enspec}
\kappa_{n,0}^{2n} = 
\frac{\int_{\scriptsize 2\pi/L}^{\infty}k^{2n}(|\hat{\bu}|^{2}+\tau^{2}|\hat{\bdf}|^{2})\,dV_{k}}
{\int_{\scriptsize 2\pi/L}^{\infty}(|\hat{\bu}|^{2}+\tau^{2}|\hat{\bdf}|^{2})\,dV_{k}}\,.
\ee
More relevant to the two-dimensional case, $\kappa_{n,1}^{2(n-1)}$ behaves as the 
$2(n-1)$th-moment of the enstrophy spectrum. Using Landau's argument the dimension of the 
global attractor $d_{L}(\mathcal{A})$ was identified with the number of degrees of freedom 
$\mathcal{N}$. In \cite{G96} a definition was introduced 
to represent the number of degrees of freedom associated with all higher derivatives of the 
velocity field represented by $\kappa_{n,r}$, which is itself an inverse length. This naturally 
leads to the definition of the infinite set
\bel{Ndef}
\mathcal{N}_{n,r} = L^{2}\left<\kappa_{n,r}^2\right>\,.
\ee
Using the definition of the quantities $\Lambda_{n,0}$ and $\Lambda_{n,1}$ ($n\geq 2$)
\bel{Lamdef1}
\Lambda_{n,0} = \frac{3n-2}{2n}\,,
\hspace{2cm}
\Lambda_{n,1} = \frac{3n-4}{2(n-1)}\,,
\ee
the second main result of the paper is a theorem whose proof is given in \S\ref{timeavs}\,:
\begin{theorem}\label{theorem2}
Let $\kappa_{n,r}$ be the moments of a two-dimensional Navier-Stokes velocity field 
defined in (\ref{2nmom1}).  Then in a two-dimensional periodic box of side $L$ the 
numbers of degrees of freedom $\mathcal{N}_{n,1}$ and $\mathcal{N}_{n,0}$ defined 
in (\ref{Ndef}) are estimated as ($n\geq 2$)
\bel{Nest1}
\mathcal{N}_{n,1} \leq c_{n,1}(a_{\ell}^{2}\Rey)^{\Lambda_{n,1}}\left(1 + \ln\Rey\right)^{1/2}\,,
\ee
\bel{Nest0}
\mathcal{N}_{n,0} \leq c_{n,0}(a_{\ell}^{2}\Rey)^{\Lambda_{n,0}}\left(1 + \ln\Rey\right)^{1/2}
\ee
where $\Lambda_{n,0}$ and $\Lambda_{n,1}$ are defined in (\ref{Lamdef1}). 
\end{theorem}
\par\noindent
Note that $\Lambda_{2,0} = \Lambda_{2,1} = 1$. Thus the estimate for the first in each sequence, 
$\mathcal{N}_{2,1}$ and $\mathcal{N}_{1,0}$, are of the same order as the estimate for $d_{L}$, 
namely $a_{\ell}^{2}\Rey (1 + \ln \Rey)^{1/3}$ except in the exponent of the logarithm. The 
exponents in (\ref{Nest1}) and (\ref{Nest0}) provide an estimate of the extra resolution that 
is needed to take account of energy at sub-Kraichnan scales. Notice that in the limit $n\to\infty$ 
both exponents converge to $3/2$.
\par\smallskip
The intermittency results of \S\ref{inter} show that there can exist small intervals of time where 
there are large \textit{lower} bounds on $\kappa_{n,1}^2$ that are much larger than the upper bound
on the long-time average for $\left<\kappa_{n,1}^2\right>$. Translated into pictorial terms, 
Figure 1 in \S\ref{inter} is consistent with the existence of spiky data whose duration must be 
very short. Estimates are found for the width of these spikes which turn out to be in terms of a 
negative exponent of $\Rey$. 


\section{\large Time average estimates in terms of $\Rey$}\label{2Dest}

\subsection{Proof of Theorem \ref{theorem1}}\label{theorem1proof}

The first step in the proof of Theorem \ref{theorem1}, which has been expressed in 
\S\ref{summary}, is to find an upper bound on $\left<H_{2}\right>$ in terms of $\Rey$. 
Consider the equation for the two-dimensional Navier-Stokes vorticity $\bw = 
\omega\,\hat{\bk}$
\bel{ns2d1}
\frac{\partial\bw}{\partial t} + \bu\cdot\nabla\bw = \nu\Delta\bw + \hbox{curl}\,\f\,,
\ee
and let $H_{n}$ be defined by $(n\geq 0)$
\bel{Hndef}
H_{n} = \int_{\Omega}|\nabla^{n}\bu|^{2}\,dV\,.
\ee
For a periodic, divergence-free velocity field $\bu$
\bel{H1def}
H_{1} = \int_{\Omega}|\nabla\bu|^{2}\,dV = \int_{\Omega}|\bw|^{2}\,dV\,.
\ee
Then the evolution equation for $H_1$ is 
\beq{en1a}
\shalf \dot{H}_{1} &=& -\nu H_{2} + \int_{\Omega}\bw\cdot\hbox{curl}\f\,dV\\
&\leq& -\nu H_{2} + \|\bu\|_{2}\|\nabla^2\f\|_{2}\label{en1b}\\
&\leq& -\nu H_{2} + \ell^{-2}\|\bu\|_{2}\|\f\|_{2}\label{en1c}\,,
\eeq
where the forcing term has been integrated by parts in (\ref{en1a}) and the narrow-band 
property has been used to move from (\ref{en1b}) to (\ref{en1c}). Using the definitions 
of $\Rey$, $\bG$, and $a_{\ell}$ in (\ref{Redef}), (\ref{Grdef}) and (\ref{aldef}), the 
long-time average of $H_{2}$ is estimated as
\beq{en2}
\left<H_{2}\right> &\leq& L^{2}\ell^{-6}\nu^{2}\Rey\,\bG\\
&\leq& c\,a_{\ell}^{2}\ell^{-4}\nu^{2}\,\Rey^{3}+ O(\Rey^{2})\,.
\eeq
This holds the key to the three results in Theorem \ref{theorem1}. 
\par\smallskip
The inverse Kraichnan length $\eta_{k}^{-6} = \epsilon_{ens}/\nu^{3}$ with
$\epsilon_{ens} = \nu L^{-2}\left<H_{2}\right>$, can now be estimated by noting that 
\bel{lemprf1}
L^{6}\epsilon_{ens} \leq c\,a_{\ell}^{6}\nu^{3}\,\Rey^{3}\,
\ee
and so
\bel{lemprf2}
L\eta_{k}^{-1} \leq c\,(a_{\ell}^{2}\Rey)^{1/2}\,,
\ee
which is (\ref{en4b}) of Theorem (\ref{theorem1}). The estimate for $d_{L}$ in (\ref{en4a}) 
then follows immediately from the relation between the estimate for $d_{L}$ in 
(\ref{in1c}) and (\ref{lemprf2}).
\par\smallskip
Finally, we turn to proving the estimate for $\left<H_{1}\right>$ in (\ref{en4c}) which turns 
around the use of the simple inequality $H_{1}^2\leq H_{2}H_{0}$. The next step is to use the 
fact that
\beq{en3a}
\left<H_{1}\right> &\leq& \left<H_{2}\right>^{1/2}\left<H_{0}\right>^{1/2}\\
&=& \nu a_{\ell}\Rey \left<H_{2}\right>^{1/2}\,.
\eeq
Using the upper bound in (\ref{en2}) gives 
\bel{en3b}
\left<H_{1}\right> \leq  c\,\nu^{2}a_{\ell}^{2}\ell^{2}\Rey^{5/2}\,,
\ee
which then gives (\ref{en4c}) in Theorem (\ref{theorem1}).  In fact, (\ref{en3b}) is an 
improvement in the bound for $\left<H_{1}\right>$ from $\Rey^{3}$ to $\Rey^{5/2}$. This result 
has also been found recently by Alexakis and Doering \cite{Alex}.

\subsection{Proof of Theorem \ref{theorem2}}\label{timeavs}

Having introduced the notation for $H_{n}$ in (\ref{Hndef}), similar quantities are used 
that contain the forcing \cite{DG02,GD05}, namely 
\bel{e:fn}
F_n = \int_{\Omega}\left(|\nabla^n \bu |^2 + \tau^2 |\nabla^n \bdf |^2\right)dV\,,
\ee
defined first in (\ref{Fdef1a}), and the moments $\kappa_{n,r}$ defined in (\ref{2nmom1})
\bel{e:kappa}
\kappa_{n,r}(t) := \left( \frac{F_n}{F_r} \right)^{\frac{1}{2 (n - r)}}.
\ee
The parameter $\tau$ in (\ref{e:fn}) is a time scale and needs to be chosen appropriately. 
The idea is that it should be chosen in such a way that the forcing does not dominate the 
behavior of the moments of the velocity field. Defining $\omega_{0} = \ell^{-2}\nu$, it 
is shown in Appendix \ref{App1.1} that this end is achieved if $\tau^{-1}$ is chosen as
\beq{taudef1}
\tau^{-1} &=& \omega_{0}[\bG(1+\ln\bG)]^{1/2}\\ 
&\leq& c\,\omega_{0}\Rey(1+\ln\Rey)^{1/2}\,.
\eeq
\par\smallskip\noindent
As a preliminary to the proof of Theorem \ref{theorem2}, we state the ladder theorem 
proved in \cite{DG02,GD05}.
\begin{theorem}\label{ladder}
The $F_n$ satisfy the differential inequalities
\begin{eqnarray}
\shalf \dot{F}_0 & \leq & - \nu F_1 + c\,\tau^{-1}F_{0}\,,
\label{e:ladder_1}\\
\shalf \dot{F}_1 & \leq & - \nu F_2 + c\,\tau^{-1}F_{1}\,,
\label{e:ladder_2}
\end{eqnarray}
and, for $n \geq 2$, either 
\bel{e:ladder_3}
\shalf \dot{F}_n \leq  - \nu F_{n+1} + c_{n,1}\Big( \|\nabla\bu\|_{\infty}
+ \tau^{-1}\Big) F_{n}\,,
\ee
\vspace{-2mm}
or
\bel{e:ladder_4}
\shalf \dot{F}_n  \leq  - \shalf\nu F_{n+1} + c_{n,2} \Big( \nu^{-1} 
\|\bu \|^2_{\infty}+ \tau^{-1}\Big) F_{n}\,.
\ee
\end{theorem}
\par\smallskip\noindent
The $L^{\infty}$-inequalities in Theorem \ref{ladder}, particularly $\|\nabla\bu\|_{\infty}$ 
in (\ref{e:ladder_3}), can be handled using a modified form of the $L^\infty$-inequality of 
Brezis and Gallouet that has already been proved in \cite{DGbook}\,:
\begin{lemma}\label{BGlem}
In terms of the $F_{n}$ of (\ref{e:fn}) and $\kappa_{3,2}$ of (\ref{e:kappa}), a modified  form 
of the two-dimensional $L^\infty$-inequality of Brezis and Gallouet is
\bel{fs0}
\|\nabla\bu\|_{\infty} \leq c\,F_{2}^{1/2}\left[1 + \ln(L \kappa_{3.2})\right]^{1/2}\,. 
\ee
\end{lemma}
\par\smallskip\noindent
This lemma directly leads to an estimate for $\left<\kappa_{n,r}^{2}\right>$ for $r\geq 2$.
\begin{lemma}\label{lem2.2}
For $n > r\geq 2$, to leading order in $\Rey$\,,
\bel{knr1}
L^{2}\left<\kappa_{n,r}^{2}\right> \leq c\,(a_{\ell}^{2}\Rey)^{3/2}(1+\ln\Rey)^{1/2}\,.
\ee
\end{lemma}
\par\smallskip\noindent
\textbf{Proof:} By dividing (\ref{e:ladder_3}) by $F_{n}$ and time averaging, we have
\bel{en5a}
\nu\left<\kappa_{n+1,n}^{2}\right> \leq c_{n,1}\left<\|\nabla\bu\|_{\infty}\right> 
+ c\,\omega_{0}\Rey(1+\ln \Rey)^{1/2}\,.
\ee
However, because $\kappa_{n,r}\leq\kappa_{n+1,n}$ for $r<n$, for every $ 2\leq r < n$, in 
combination with Lemma \ref{BGlem}, we have 
\bel{en5b}
\nu\left<\kappa_{n,r}^{2}\right> \leq c\,\left<F_{2}^{1/2}
\left[1 + \ln (L\kappa_{3,2})\right]^{1/2}\right> + c\,\omega_{0}\Rey(1+\ln \Rey)^{1/2}\,.
\ee
The logarithm is a concave function and $\kappa_{3,2} \leq \kappa_{n,r}$ so Jensen's inequality gives
\bel{en5c}
L^{2}\left<\kappa_{n,r}^{2}\right> \leq L^{2}\nu^{-1}c\,\left<F_{2}\right>^{1/2}
\left<\left[1 + \ln \{L^{2}\left<\kappa_{n,r}^2\right>\}\right]\right>^{1/2} 
+   c\,a_{\ell}^{2}\Rey(1+\ln \Rey)^{1/2}\,.
\ee
The estimate for $\left<F_{2}\right>$ can be found from $\left<H_{2}\right>$ in (\ref{en2}); the 
extra term $\tau^{2}\|\nabla^2\bdf\|_{2}^{2}$ is no more than $O(\Rey^2)$.  Standard properties 
of the logarithm turn inequality (\ref{en5c}) into (\ref{knr1}). \hspace{2cm}$\blacksquare$
\par\smallskip
Lemma \ref{lem2.2} gives estimates for $\left<\kappa_{n,r}^{2}\right>$ for $r\geq 2$. These 
are used in the following theorem to give better estimates for the cases $r=0$ and $r=1$. Prior 
to this, it is necessary to state the results that immediately derive from (\ref{e:ladder_1}) 
and (\ref{e:ladder_2}) by respectively dividing through by $F_{0}$ and $F_{1}$ before time 
averaging
\bel{thm3c}
\mathcal{N}_{1,0} \equiv L^{2}\left<\kappa_{1,0}^{2}\right>\leq  c\,a_{\ell}^{2}\Rey(1+\ln \Rey)^{1/2}\,,
\hspace{1cm}
\mathcal{N}_{2,1} \equiv L^{2}\left<\kappa_{2,1}^{2}\right> 
\leq  c\,a_{\ell}^{2}\Rey(1+\ln \Rey)^{1/2}\,.
\ee
With the estimates in (\ref{thm3c}) we are now ready to complete the proof of Theorem \ref{theorem2}.
\par\smallskip\noindent
\textbf{Proof of Theorem \ref{theorem2}:} Let us return to (\ref{knr1}) in Lemma \ref{lem2.2} and 
use the fact that
\bel{kn1}
\left<\kappa_{n,1}^{2}\right> = 
\left<\left(\frac{F_{n}}{F_{2}}\right)^{\frac{1}{n-1}}
\left(\frac{F_{2}}{F_{1}}\right)^{\frac{1}{n-1}}\right>
= \left<\kappa_{n,2}^{\frac{2(n-2)}{n-1}}\kappa_{2,1}^{\frac{2}{n-1}}\right> \,,
\ee
and thus
\bel{kn2}
\left<\kappa_{n,1}^{2}\right> \leq 
\left<\kappa_{n,2}^{2}\right>^{\frac{n-2}{n-1}}\left<\kappa_{2,1}^{2}\right>^{\frac{1}{n-1}}\,.
\ee
Using (\ref{knr1}) in Lemma \ref{lem2.2}, together with (\ref{thm3c}), for $n \geq 2$,
\bel{kn4}
\mathcal{N}_{n,1}  = L^{2}\left<\kappa_{n,1}^{2}\right> \leq c_{n,1}
\,(a_{\ell}^{2}\Rey)^{\frac{3n-4}{2(n-1)}}
\left[1 + \ln \Rey\right]^{1/2}\,,
\ee
which coincides with $a_{\ell}^{2}\Rey(1+\ln\Rey)^{1/2}$ at $n=2$ but converges to 
$\Rey^{3/2}(1+\ln\Rey)^{1/2}$ as $n\to\infty$.  The exponent $\Lambda_{n,1}$ is defined 
in (\ref{Lamdef1}).
\par\smallskip
Likewise, in the same manner as (\ref{kn1}) we have
\bel{kn5}
\left<\kappa_{n,0}^{2}\right> \leq
\left<\kappa_{n,1}^2\right>^{\frac{n-1}{n}}\left<\kappa_{1,0}^2\right>^{\frac{1}{n}}\,.
\ee
Thus we find that for $n \geq 2$
\bel{kn6}
\mathcal{N}_{n,0}  = L^{2}\left<\kappa_{n,0}^{2}\right> \leq c_{n,0}\,(a_{\ell}^{2}\Rey)^{\frac{3n-2}{2n}}
\left[1 + \ln \Rey\right]^{1/2}\,. 
\ee
The exponent $\Lambda_{n,0}$ is defined in (\ref{Lamdef1}).\hspace{8cm}$\blacksquare$

%
%
%
\section{\large Point-wise Estimates}\label{point}

Let us consider the differential inequalities for $H_{0}$ and $H_{1}$:
\bel{H0}
\shalf \dot{H}_{0} \leq -\nu H_{1} + \|\bdf\|_{2}H_{0}^{1/2}\,,
\ee
\bel{H1}
\shalf \dot{H}_{1} \leq -\nu H_{2} + \ell^{-2}\|\bdf\|_{2}H_{0}^{1/2}\,,
\ee
having used the narrow-band property on (\ref{H1}). Upon combining Poincar\'{e}'s inequality
with Lemmas \ref{applem1} and \ref{applem2} in Appendix \ref{App1.2} we obtain
\bel{H01}
\overline{\lim}_{t\to\infty}H_{0} \leq c\,a_{\ell}^{6}\nu^{2}\bG^{2} 
\leq c\,a_{\ell}^{6}\nu^{2}\Rey^{4}\,,
\ee
and
\bel{H11}
\overline{\lim}_{t\to\infty}H_{1} \leq c\,\ell^{-2}a_{\ell}^{6}\nu^{2}\bG^{2} 
\leq c\,\ell^{-2}a_{\ell}^{6}\nu^{2}\Rey^{4}\,.
\ee
The additive forcing terms in $F_1$ and $F_0$ are of a lower order in $\Rey$ so we end up with 
\bel{H04}
\overline{\lim}_{t\to\infty}F_{0}\leq c\,a_{\ell}^{6}\nu^{2}\Rey^{4} + O(\Rey^2)\,,
\ee
\bel{H14}
\overline{\lim}_{t\to\infty}F_{1} \leq c\,\ell^{-2}a_{\ell}^{6}\nu^{2}\Rey^{4} + O(\Rey^2)\,.
\ee
The estimate for  $F_1$ enables us to obtain point-wise estimates on 
$F_n, \, n \geq 2$ \cite[sec. 7.2]{DGbook}. In fact we have the following lemma.
\begin{lemma}\label{Fnlem}
As $Gr \rightarrow \infty$
\begin{equation}
\overline{\lim}_{t \rightarrow \infty} F_n \leq c_{n}\nu^{2}\ell^{-2n}a_{\ell}^{6n}\Rey^{4n}\,.
\label{e:fn_est}
\end{equation}
\end{lemma}
\par\smallskip\noindent
\textbf{Proof:}  Applying a Gagliardo--Nirenberg inequality in two-dimensions to $\nabla\bu$ we obtain
\bel{e:gn}
\| \nabla \bu \|_{\infty} \leq c\, \| \nabla^n \bu \|_{2}^{a} \| \nabla \bu\|_{2}^{1 - a} 
\leq c F_n^{\frac{a}{2}} F_1^{\frac{1-a}{2}}\,,
\ee
with $a = \frac{1}{n - 1}$. Using this in \eqref{e:ladder_3} gives
\bel{extra1}
\shalf \dot{F}_n \leq - \nu F_{n+1} + c_{n}F_n^{1 +\frac{a}{2}} F_1^{\frac{1 - a}{2}} + 
c\,\omega_{0}\Rey(1+\ln\Rey)^{1/2} F_{n}\,.
\ee
Moreover the following inequality can easily be proved using Fourier transforms
\bel{FNinequal}
F_N^{p +q} \leq F_{N-p}^q F_p^{N + q}\,,
\ee
from which, with $N = n, \, p = n - 1, \, q =1$, it can be deduced that
\begin{equation}\label{e:interp}
- F_{n +1} \leq - \frac{F_n^{\frac{n}{n - 1}}}{F_1^{\frac{1}{n - 1}}}\,.
\end{equation}
We now use \eqref{e:interp} in \eqref{extra1} to obtain
\begin{equation}\label{e:fn_ineq}
\shalf \dot{F}_n \leq - \nu \frac{F_n^{\frac{n}{n -1}}}{F_1^{\frac{1}{n - 1}}}
+ c_{n}F_n^{1 +\frac{a}{2}} F_1^{\frac{1 - a}{2}} + c\,\omega_{0}\Rey(1+\ln\Rey)^{1/2} F_n,
\end{equation}
with $a =\frac{1}{n - 1}$. We use now estimate (\ref{H14}) in \eqref{e:fn_ineq} with the 
further use of Lemma \ref{applem2} to obtain 
\bel{kappalem2}
\overline{\lim}_{t \rightarrow \infty} F_n \leq c_{n}\nu^{2}\ell^{-2n}a_{\ell}^{6n}\bG^{2n}\,,
\ee
which leads to the result.\hspace{10cm} $\blacksquare$
\par\bigskip\noindent
The above Lemma enables us to obtain an estimate on the wave-numbers $\kappa_{n,r}$.
\begin{lemma}\label{kappalem1}
For $n > r \geq 0$, as $\bG \rightarrow \infty$
\begin{equation}\label{kappalim}
\overline{\lim}_{t \rightarrow \infty} (L \kappa_{n,r}) 
\leq c_{n} a_{\ell}^{\frac{4n-r-1}{n-r}}
\Rey^{\frac{2n -1}{n - r}}(1+\ln\Rey)^{\frac{1}{2(n-r)}}\,.
\end{equation}
\end{lemma}
\par\smallskip\noindent
\textbf{Proof:} Essentially one uses the upper bound on $F_n$ and the lower bound on $F_r$ 
which can be calculated from the forcing part in terms of $\bG$, leading to the result
(see also \cite[Ch. 7]{DGbook}). $\blacksquare$


\section{\large Intermittency: good and bad intervals}\label{inter}

The issue of intermittency in solutions of the two-dimensional Navier-Stokes equations is now 
addressed. While the $F_{n}$ and $\kappa_{n,r}$ are bounded from above for all time, nevertheless 
it is possible that their behaviour could be spiky in an erratic manner. To show how this might 
come about, consider the definition of $\kappa_{n,r}$ in (\ref{2nmom1}) from which we find 
\begin{equation}\label{gb1}
F_{n +1} = \kappa_{n,r}^{2}\left(\frac{\kappa_{n+1,r}}{\kappa_{n,r}}\right)^{2(n +1 - r)} F_{n}\,.
\end{equation}
Now consider inequality (\ref{extra1}) re-written as
\bel{gb3}
\shalf\frac{\dot{F}_{n}}{F_{n}} 
\leq - \nu \kappa_{n,1}^{2}\left(\frac{\kappa_{n+1,1}}{\kappa_{n,1}}\right)^{2n} 
+ c_{n}\left(\frac{\kappa_{n+1,1}}{\kappa_{n,1}}\right)^{n}\kappa_{n,1}F_{1}^{1/2} 
+ c\,\omega_{0}\Rey(1+\ln\Rey)^{1/2}\,.
\ee
where we have used (\ref{gb1}) and the fact that $\kappa_{n,1}\leq \kappa_{n+1,1}$ in the middle 
term. Using Young's inequality on this same term we end up with 
\bel{gb4}
\shalf\frac{\dot{F}_{n}}{F_{n}} \leq 
- \shalf \nu \kappa_{n,1}^{2}\left(\frac{\kappa_{n+1,1}}{\kappa_{n,1}}\right)^{2n} 
+ c_{n}\nu^{-1}F_{1} + c\,\omega_{0}\Rey(1+\ln\Rey)^{1/2}\,.
\ee
The main question is whether, for Navier-Stokes solutions, the lower bound on 
\bel{gb5}
\frac{\kappa_{n+1,1}}{\kappa_{n,1}} \geq 1
\ee
can be raised from unity. A variation on the interval theorem proved in \cite{GD05} is used.
\begin{theorem}\label{intervalthm}
For any value of the parameter $\mu \in (0,1)$, the ratio $\kappa_{n+1,1}/\kappa_{n,1}$ obeys 
the long-time averaged inequality $(n\geq 2)$
\bel{as4}
\left<\left[c_{n}\left(\frac{\kappa_{n+1,1}}{\kappa_{n,1}}\right)^2\right]^{1/\mu -1}
-\left[\frac{(L^2\kappa_{n,1}^{2})^{\mu}}{(a_{\ell}^{2}\Rey)^{\Lambda_{n,1}}
(1+\ln\Rey)^{1/2}}\right]^{1/\mu -1}\right>\geq 0\,,
\ee
where the $c_{n}$ are the same as those in Theorem (\ref{theorem2}).  Hence there exists at 
least one interval of time, designated as a `good interval', on which the inequality
\bel{thm1gb}
c_{n}\left(\frac{\kappa_{n+1,1}}{\kappa_{n,1}}\right)^2 
\geq \frac{(L^2\kappa_{n,1}^{2})^{\mu}}{(a_{\ell}^{2}\Rey)^{\Lambda_{n,1}}(1+\ln\Rey)^{1/2}}
\ee
holds. Those other parts of the time-axis on which the reverse inequality 
\bel{thm2gb}
c_{n}\left(\frac{\kappa_{n+1,1}}{\kappa_{n}}\right)^2 < 
\frac{(L^2\kappa_{n,1}^{2})^{\mu}}{(a_{\ell}^{2}\Rey)^{\Lambda_{n,1}}(1+\ln\Rey)^{1/2}}
\ee
holds are designated as 'bad intervals'.
\end{theorem}
\par\smallskip\noindent
\textbf{Remark:} In principle, the whole time-axis could be a good interval, 
whereas the positive time average in (\ref{as4}) ensures that the complete 
time-axis cannot be `bad'. This paper is based on the worst-case supposition 
that bad intervals exist, that they could be multiple in number, and that the 
good and the bad are interspersed. The precise distribution and occurrence of 
the good/bad intervals and how they depend on $n$ remains an open question. 
The contrast between the two-dimensional and three-dimensional Navier-Stokes 
equations is prominent; while no singularities can occur in the $\kappa_{n,1}$ 
in the two-dimensional case, in three dimensions it is within these bad intervals 
that they can potentially occur.
\par\smallskip\noindent
\textbf{Proof:} Take two parameters $0 < \mu < 1$ and $0 < \alpha < 1$ such 
that $\mu + \alpha = 1$. The inverses $\mu^{-1}$ and $\alpha^{-1}$ will be 
used as exponents in the H\"{o}lder inequality on the far right hand side of
\bel{as2}
\left<\kappa_{n,1}^{2\alpha}\right> \leq
\left<\kappa_{n+1,1}^{2\alpha}\right> = \left<\left(\frac{\kappa_{n+1,1}}
{\kappa_{n,1}}\right)^{2\alpha}\kappa_{n,1}^{2\alpha}\right> 
\leq \left<\left(\frac{\kappa_{n+1,1}}{\kappa_{n,1}}\right)^{2\alpha/\mu}
\right>^{\mu}\left<\kappa_{n,1}^2\right>^{\alpha}\,,
\ee
thereby giving 
\bel{as3}
\left<\left(\frac{\kappa_{n+1,1}}{\kappa_{n,1}}\right)^{2\alpha/\mu}\right>
\geq \left(\frac{\left<\kappa_{n,1}^{2\alpha}\right>}
{\left<\kappa_{n,1}^2\right>^{\alpha}}\right)^{1/\mu} 
= \left<\kappa_{n,1}^{2\alpha}\right> \left(\frac{\left<\kappa_{n,1}^{2\alpha}\right>}
{\left<\kappa_{n,1}^2\right>}\right)^{\alpha/\mu}.
\ee
Two-dimensional Navier-Stokes information can be injected into these formal manipulations: 
the upper bound on $\left<\kappa_{n,1}^2\right>$ from Theorem (\ref{theorem2}) and the lower 
bound $L\kappa_{n,1}\geq 1$ are used in the ratio on the far right hand side of 
(\ref{as3}) to give (\ref{as4}), with the same $c_{n}$ as in Theorem 
(\ref{theorem2}).~~~~$\blacksquare$
\par\smallskip
Now consider what must happen on bad intervals. It is always true that 
$\kappa_{n+1,1}/\kappa_{n,1} \geq 1$, so (\ref{thm2gb}) implies that on these intervals 
there is a lower bound 
\bel{gb6a}
L^2\kappa_{n,1}^{2} > c_{n}(a_{\ell}^{2}\Rey)^{\Lambda_{n,1}/\mu}(1+\ln\Rey)^{1/2\mu}\,.
\ee
This lower bound cannot be greater than the upper point-wise bound in (\ref{kappalim}), 
which means that $\mu$ is restricted by
\bel{gb6b}
\frac{\Lambda_{n,1}}{\mu} < 2\left(\frac{2n-1}{n-1}\right)\,.
\ee
Moreover, the factor of $1/\mu$ in the exponent makes the lower bound in (\ref{gb6a}) much 
larger than the \textit{upper} bound on the average $\left<\kappa_{n,1}^{2}\right>$ given 
in Theorem (\ref{theorem2}). \textit{These intervals must therefore be very short}. To 
estimate how large they can be requires an integration of (\ref{gb4}) over short times 
$\Delta t = t-t_{0}$ which, in turn, requires the time-integral of $H_{1}$ for short 
times $\Delta t$. We use the notation $\Idt = \It$, with the definition $\omega_{0}=\nu\ell^{-2}$.
\begin{lemma}\label{deltalem}
To leading order in $\Rey$
\bel{H03}
\Idt F_{1}\,dt\leq 
\nu a_{\ell}^{4}\left[c_{1}a_{\ell}^{2}+ c_{2}\,\omega_{0}\Delta t\right]\Rey^{4}\,.
\ee
\end{lemma}
\par\smallskip\noindent
\textbf{Proof:} Integrating (\ref{H0}) over a short time $\Delta t$ gives
\beq{H02}
\nu \Idt H_{1}\,dt &\leq & \shalf H_{0}(t_{0}) 
+ \Delta t\left[\ell^{-2}\nu^{3}a_{\ell}^{4}\bG^{2}\right]\nonumber\\
&\leq & c_{1}a_{\ell}^{6}\nu^{2}\Rey^{4} 
+ \Delta t\left[c_{2}\ell^{-2}\nu^{3}a_{\ell}^{4}\Rey^{4}\right]\,,
\eeq
having used (\ref{H01}) for the $\shalf H_{0}(t_{0})$-term. The forcing term in $F_{1}$ is 
only $O(\Rey^2)$. \hspace{2cm}$\blacksquare$
\par\vspace{-1.5cm}
$$
\begin{minipage}[ht]{9cm}
\setlength{\unitlength}{.75cm}
\begin{picture}(11,11)(0,0)
\thicklines
\put(0,0){\vector(0,1){8}}
\put(0,0){\vector(1,0){10}}
\put(0,8.2){\makebox(0,0)[b]{$\kappa_{n,1}(t)$}}
\put(10.2,0){\makebox(0,0)[b]{$t$}}
\thinlines
\put(3,0){\line(0,1){8}}
\put(3.5,0){\line(0,1){8}}
\put(7,0){\line(0,1){8}}
\put(7.5,0){\line(0,1){8}}
\multiput(0,2)(.1,0){90}{.}
\multiput(0,6)(.1,0){30}{.}
\multiput(3.5,6)(.1,0){55}{.}
\put(-.8,2){\makebox(0,0)[b]{\scriptsize$Re^{\Lambda_{n,1}}$}}
\put(5.3,2.1){\makebox(0,0)[b]{\scriptsize Long-time average}}
\put(5.1,6.4){\makebox(0,0)[b]{\scriptsize$L\kappa_{n,1}>Re^{\Lambda_{n,1}/\mu}$}}
\put(3.9,6.4){\vector(-1,0){.72}}
\put(6.2,6.4){\vector(1,0){1.1}}
\put(-.8,6){\makebox(0,0)[b]{\scriptsize$Re^{\Lambda_{n,1}/\mu}$}}
\put(-.85,7.1){\makebox(0,0)[b]{\scriptsize$\max(\kappa_{n,1})$}}
\multiput(0,7)(.1,0){90}{.}
\put(3.25,-.70){\makebox(0,0)[b]{\scriptsize$(\Delta t)_{b}$}}
\put(5.25,-.68){\makebox(0,0)[b]{\scriptsize$(\Delta t)_{g}$}}
\put(5.75,-.25){\vector(1,0){1.25}}
\put(4.75,-.25){\vector(-1,0){1.2}}
\qbezier[250](0,1)(3,0)(3.01,7.70)
\qbezier(3.5,7.0)(5,-6)(7,7.0)
\qbezier[300](7.5,7.70)(7.5,0)(10,1)
\end{picture}
\end{minipage}
$$
\par\vspace{.75cm}\noindent
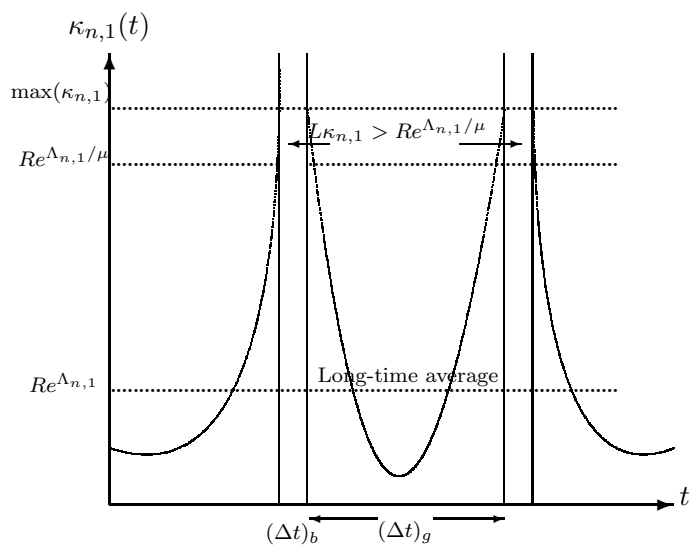
{\textbf{Figure 1:} {\small A cartoon, not to scale, of 
good/bad intervals for some value of $n\geq 3$.}\label{knfig}}
\par\vspace{.5cm}\noindent
Now we wish to estimate $\omega_{0}\Delta t$ in terms of $\Rey$. Integrating (\ref{gb4}), 
using (\ref{H02}) and the lower bound (\ref{gb6a}) and multiplying by $\ell^{2}$, we have
\beq{gb7}
\shalf\ell^{2}\left[\ln F_{n}(t) - \ln F_{n}(t_{0})\right] &+& \shalf c_{n}\nu a_{\ell}^{-2}
(a_{\ell}^{2}\Rey)^{\Lambda_{n,1}/\mu}(1+\ln\Rey)^{1/2\mu}\Delta t\nonumber\\
&\leq & \ell^{2} a_{\ell}^{4}\left[c_{1}a_{\ell}^{2}+ c_{2}\,\omega_{0}\Delta t\right]
\Rey^{4}\nonumber\\
&+& c\,\ell^{2}\omega_{0}\Delta t\Rey(1+\ln\Rey)^{1/2}\,.
\eeq
As $\bG\to\infty$, the dominant terms are 
\bel{gb8}
\omega_{0}\Delta t\left\{a_{\ell}^{-2}(a_{\ell}^{2}\Rey)^{\Lambda_{n,1}/\mu}(1+\ln\Rey)^{1/2\mu}
- a_{\ell}^{6}\Rey^{4}\right\}\leq c_{1}a_{\ell}^{6}\Rey^{4}\,.
\ee
Choosing $\mu$ in the range, to leading order we have
\bel{gb9}
\mu < \squart \Lambda_{n,1}\,,
\ee
then $\Delta t$ must satisfy
\bel{gb10}
\omega_{0}\Delta t \leq c\,(a_{\ell}^{2}\Rey)^{4- \Lambda_{n,1}/\mu}\,.
\ee
Because the exponent in (\ref{gb10}) is necessarily negative these intervals are very small 
and decreasing with increasing $\Rey$. Combining (\ref{gb6b}) with (\ref{gb9}) we have
\bel{gb12}
\frac{(n-1)}{2(2n-1)}\Lambda_{n,1} < \mu < \squart \Lambda_{n,1}\,,
\ee
which actually holds for every $n\geq 1$. Figure 1 is a cartoon-like figure displaying 
the lower bound on the bad intervals of width $(\Delta t)_{b}$ and also the maximum 
of $\kappa_{n,1}$ allowed by (\ref{kappalim}) in Lemma \ref{kappalem1}. The full dynamics 
of two-dimensional Navier-Stokes is actually determined by the intersection of all 
cartoons for every $n\geq 3$ on the grounds that the position and occurrence of the 
bad intervals varies with $n$. Thus we are interested in the limit $n\to\infty$ 
which determines that the range of $\mu$ is squeezed between
\bel{gb13}
\frac{3}{8}\left(1-\frac{5}{6n}\right) < \mu < \frac{3}{8}\left(1-\frac{1}{3n}\right)\,.
\ee
Thus, in the limit, $\mu$ takes a value just under $3/8$. We conclude that the interval 
theorem (Theorem \ref{intervalthm}) reproduces the effects of 
intermittency in a two-dimensional flow by manifesting very large lower bounds within bad 
intervals and suppressing spiky behaviour within the good intervals which must be quiescent 
for long intervals, otherwise the long-time average would be violated.

\par\vspace{0.25cm}\noindent
\textbf{Acknowledgements:} The authors would like to thank Matania Ben-Artzi, Charles Doering, 
Darryl Holm, Haggai Katriel and Edriss Titi for comments \& suggestions. JDG would also 
like to thank the Mathematics Departments of the Weizmann Institute of Science and the Hebrew 
University of Jerusalem for their hospitality during December 2005 and January 2006 when some 
of these ideas were conceived.

%
%
%
\appendix
\section{\large Forcing \& the fluid response}\label{App1.1}

For technical reasons, we must address the possibility that in their evolution the quantities 
$H_{n}$ might take small values. Thus we need to circumvent problems that may arise when dividing 
by these (squared) semi-norms. We follow Doering and Gibbon \cite{DG02} who introduced the modified
quantities
\bel{Fndef}
F_{n} = H_{n} + \tau^{2}\|\nabla^{n}\bdf\|^{2}_{2}\,,
\ee
where the ``time-scale'' $\tau$ is to be chosen for our convenience. So long as
$\tau \neq 0$, the $F_{n}$ are bounded away from zero by the explicit value
$\tau^{2}L^{3}\ell^{-2n}f_{rms}^2$. Moreover, we may choose $\tau$ to depend on 
the parameters of the problem such that $\left<F_{n}\right>\sim
\left<H_{n}\right>$ as $\bG\to\infty$. To see how to achieve this, let us define
\bel{taudef2}
\tau = \ell^{2}\nu^{-1}[\bG(1+\ln\bG)]^{-1/2}\,.
\ee
Then the additional term in (\ref{Fndef}) is
\begin{eqnarray}\label{add2}
\tau^{2}\|\nabla^{n}\bdf\|_{2}^{2} &=& 
L^{3}\nu^{-2}\ell^{4-2n}f_{rms}^{2}[\bG(1+\ln\bG)]^{-1}\nonumber\\
&=& \nu^{2}\ell^{-(2n+2)}L^{3}\bG(1+\ln\bG)^{-1}\,.
\end{eqnarray}
Now Doering \& Foias \cite{DF} proved that in $d$-dimensions, the energy dissipation rate 
$\epsilon$ has a lower bound of the form
\bel{epslb}
\epsilon \geq c\,\nu^{3}\ell^{-3}L^{-1}\bG\,.
\ee
Using this on the far right hand side of (\ref{add2}) we arrive at 
\begin{eqnarray}\label{add3}
\tau^{2}\|\nabla^{n}\bdf\|_{2}^{2} 
&\leq & c_{6}\epsilon\,\ell^{-(2n-1)}L^{4}\nu^{-1}(1+\ln\bG)^{-1}\nonumber\\
& = & c_{6}\left(\frac{L}{\ell}\right)^{(2n-1)}
L^{-2(n-1)}\bigl<H_{1}\bigr>(1+\ln\bG)^{-1}\,.
\end{eqnarray}
Using Poincar\'{e}'s inequality in the form 
$H_{1} \leq (2\pi L)^{2(n-1)}H_{n}$, as $\bG\to\infty$ we have
\bel{add4}
\frac{\tau^{2}\|\nabla^{n}\bdf\|_{2}^{2}}{\bigl<H_{n}\bigr>} \leq
c_{6}a_{\ell}^{(2n-1)}(1+\ln\bG)^{-1}\,.
\ee
Hence, the additional forcing term in (\ref{Fndef}) 
becomes negligible with respect to $\left<H_{n}\right>$ as $\bG\to \infty$, so the 
forcing does not dominate the response.

%
%

\section{\large Comparison theorems for ODEs}\label{App1.2}

We present a comparison theorem for ODE which is useful for obtaining various estimates. We
start with the following classical result.
\begin{lemma}\label{applem1}
Let $f:[0,T]\times \R \rightarrow \R$ be a continuous function which is locally Lipschitz 
uniformly in $t$: for all intervals $[a,b] \subset \R$ there exists a constant such that
$|f(s,x) - f(s,y)| \leq C |x - y|$ for all $x, \, y \in [a,b]$ and all $s \in [0,T]$.
Furthermore, let $x \in AC([0,T],\R)$ be such that 
$$
\dot{x}(t) \leq f(t, x(t))  
$$
for all $t \in [0,T]$ and let $y(t)$ be the solution of $\dot{y}(t) = f(t,y(t))$ on
$[0,T]$. Assume further that $x(0) \leq y(0)$. Then, $x(t) \leq y(t)$ for all $t \in
[0,T]$.
\end{lemma}
\par\smallskip\noindent
We can use this Lemma to prove the following useful result.
\begin{lemma}\label{applem2}
Let $x: [0,T] \rightarrow [0, \infty)$ be an absolutely continuous function with $x(0) > 0$
which satisfies 
\begin{equation}
\dot{x} \leq \Delta_{0} x  + F x^{n_1} - E x^{n_2},
\label{e:ineq}
\end{equation}
where $\Delta_{0}, \, F, \, E >0$ and $1 < n_1 < n_2$. Then
\begin{equation}
\limsup_{t \rightarrow \infty} x(t) \leq (4 \Delta_{0} E^{-1})^{\frac{1}{n_2 -1}} + 
(2 F E^{-1})^{\frac{1}{n_2 -n_1}}.
\label{e:ineq_est}
\end{equation}
\label{lem:ineq_est}
\end{lemma}


\bibliographystyle{unsrt}

\end{document}